\documentclass[aip,reprint,amsfonts,amsmath,amssymb,floatfix,citeautoscript,groupedaddress]{revtex4-1}
\usepackage{graphicx}% include figure files
\usepackage{dcolumn}% align table columns on decimal point
\usepackage{bm}% bold math
\usepackage{parskip}
\usepackage{color}
\usepackage{enumerate}
\usepackage{psfrag}
\usepackage{upgreek}
\setcitestyle{numbers,square}
\setcitestyle{notesep={ }}

\begin{document}

\title{
%further results on the phase behavior of hard circular arcs
non-nematicity of the filamentary phase in systems of hard minor circular arcs
}

\author{Juan Pedro Ram\'irez Gonz\'alez}
\affiliation{
Departamento de F\'isica Te\'orica de la Materia Condensada,  \\
Universidad Aut\'onoma de Madrid,
Ciudad Universitaria de Cantoblanco, \\
%Campus de Cantoblanco, \\
E-28049 Madrid, Spain
}

\author{Giorgio Cinacchi}
%\email{giorgio.cinacchi@uam.es}
\affiliation{
Departamento de F\'isica Te\'orica de la Materia Condensada,  \\
Instituto de F\'isica  de la Materia Condensada (IFIMAC),    \\
Instituto de Ciencias de Materiales ``Nicol\'{a}s Cabrera'', \\
Universidad Aut\'onoma de Madrid,
Ciudad Universitaria de Cantoblanco, \\
%Campus de Cantoblanco, \\
E-28049 Madrid, Spain
}

\date{\today}

\begin{abstract}
This work further investigates an aspect of the phase behavior of hard circular arcs,
whose phase diagram has been recently calculated by Monte Carlo numerical simulations:
the non-nematicity of the filamentary phase 
that hard minor circular arcs form.
Both second-virial density-functional theory and 
further Monte Carlo numerical simulations find 
that the positional one-particle density function is undulate
in the direction transverse to the axes of the filaments 
while
further Monte Carlo numerical simulations find 
that the mobility of the hard minor circular arcs across the filaments 
occurs via a mechanism 
%that is 
reminiscent of the mechanism of diffusion in a smectic phase:
the filamentary phase is not a \{``modulated'' [``splay(-bend)'']\} 
nematic phase.
%Consequently, it is not a ``modulated'' [``splay(-bend)''] nematic phase
%and its not being nematic questions whether such a 
%``modulated'' [``splay(-bend)''], truly nematic, 
%phase could ever exist.
\end{abstract}

\maketitle

\section{introduction}
\label{introduction}
\begin{figure}[h!]
\includegraphics[scale=0.525]{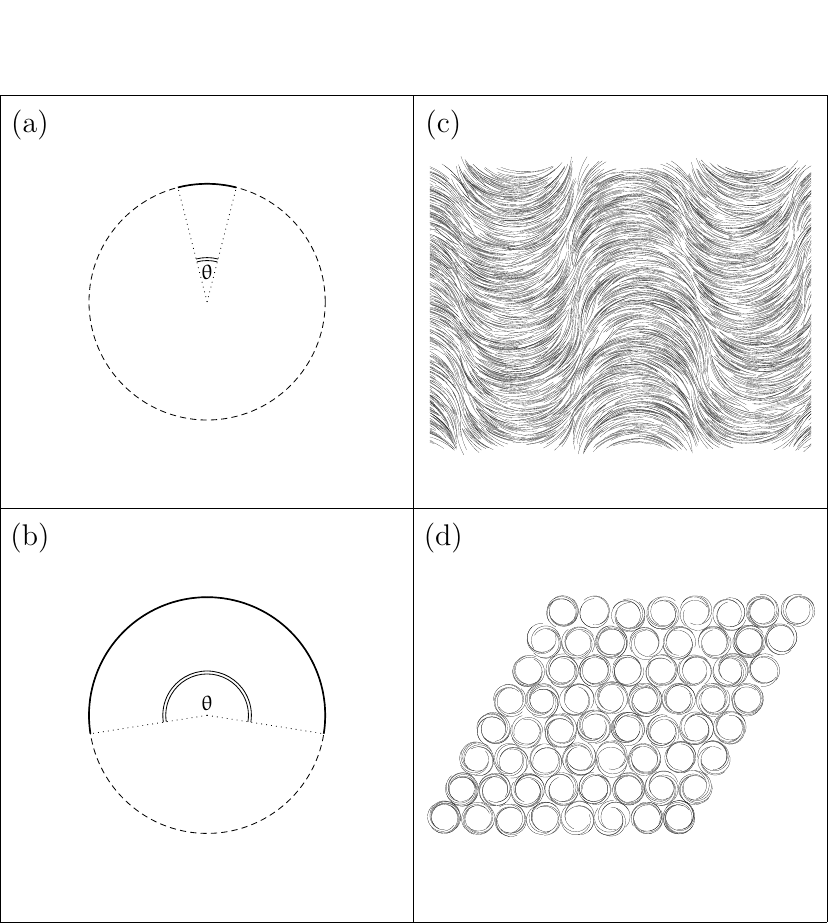}
\caption{
(a) Example of a minor circular arc with subtended angle \(\uptheta=0.5\) (continuous line) 
and its parent circumference (discontinuous line). 
(b) Example of a major circular arc with subtended angle \(\uptheta = 1.1\uppi\) (continuous line)
and its parent circumference (discontinuous line).
(c) Portion of a configuration 
of a system of hard minor circular arcs with subtended angle \(\uptheta = 0.5\) 
in the filamentary phase. 
(d) Configuration 
of a system of hard major circular arcs with subtended angle \(\uptheta= 1.1\uppi\) 
in the cluster hexagonal phase.}
\label{figura1}
\end{figure}

Two previous articles have reported results on 
the dense packings \cite{arcouno} and the phase behavior \cite{arcodue} 
of hard infinitesimally--thin circular arcs
in the two--dimensional Euclidean space \(\mathbb{R}^2\).
These two previous articles indicate 
that 
both hard minor circular arcs, 
with a subtended angle \(\uptheta \in [0,\uppi]\) 
[Fig. \ref{figura1} (a)], 
and hard major circular arcs, 
with a subtended angle \(\uptheta \in (\uppi,2\,\uppi]\) 
[Fig. \ref{figura1} (b)], 
each form, at sufficiently high density, 
a distinct (entropic) phase 
whose structural unit is \textit{supraparticular}:
hard minor circular arcs form a filamentary phase 
in which these hard curved particles
tend to organize along the parent (semi)circumference, 
so piling up into filaments, 
which in turn organize side-up by side-down 
[Fig. \ref{figura1} (c)]
\cite{arcodue};
hard major circular arcs form a cluster hexagonal phase 
in which a number of these hard curved particles 
suitably intertwine into compact roundels, 
%tightening up to the parent (circumference) circle on densification, 
which in turn organize at the sites of a triangular lattice 
[Fig. \ref{figura1} (d)]
\cite{arcodue}.
Qualitatively: hard circular arcs tend to reconstruct 
the hard particle from which they have been severed:
hard minor circular arcs tend to reconstruct 
the parent hard semicircumference and
the hard semicircumferences so reconstructed pile up into filaments
which alternately reverse while succeeding one another  
[Fig. \ref{figura1} (c)];
hard major circular arcs tend to reconstruct, 
by suitably intertwining, the parent hard circumference  and
the hard circumferences so reconstructed 
dispose themselves at the sites of a triangular lattice 
[Fig. \ref{figura1} (d)] \cite{notuccia}.

The present article reports further results on 
the first of these two cluster (entropic) phases, the filamentary phase
[Fig. \ref{figura1} (c)] \cite{arcodue},
that further indicate its non-nematicity.

In fact, the filamentary phase [Fig. \ref{figura1} (c)] \cite{arcodue} can be confounded with 
a conjectural ``modulated'' [``splay(-bend)''] nematic phase 
which is locally polar, 
with the polar director \(\hat{\mathbf p}\) 
that periodically varies along an axis \textsl{x},
\({\hat{\mathbf p}}({\rm x}) =
( \cos(\Uptheta({\rm x})), \sin(\Uptheta({\rm x})) )\),
and the nematic director \(\hat{n}\) that also periodically
varies along the same axis \textsl{x}, 
\(\hat{n}({\rm x}) = 
( \pm \cos(\Uptheta({\rm x})), \pm \sin(\Uptheta({\rm x})))\) 
[or
\(\hat{n}({\rm x}) = 
( \pm \sin(\Uptheta({\rm x})), \mp \cos(\Uptheta({\rm x})))\)],
%;
i.e., a phase in two dimensions analogue to 
%a phase in three dimensions
%maybe reminiscent of 
that conjectural ``modulated'' nematic phase in three dimensions 
that was denominated ``splay(-bend)'' and 
predicted on the basis of continuum elasticity theory considerations 
\cite{meyerdozov,splaybend}.
Yet, even leaving aside the concern that can originate from 
the comparison between the particle length scale with the periodicity length scale, 
with the former that should be significantly shorter than the latter
so that a continuum elasticity theory be applicable,
to qualify for being nematic a phase must be positionally uniform.

The non-uniformity of the filamentary phase \cite{arcodue} is further demonstrated 
by using both second-virial density-functional theory \cite{wu} results and
further Monte Carlo numerical simulation \cite{MCorigin,MCWood,MClibri} results (Section \ref{nonema}). 
The former analytical results comprise 
the, necessarily approximate, sequence of the phases, isotropic, nematic, filamentary, 
that a system of hard minor circular arcs form 
as a function of \(\uptheta\) and the number density \(\uprho\) 
along with the corresponding one-particle density functions (Section \ref{nonema1}).
The latter numerical results comprise 
the one-particle density functions (Section \ref{nonema1}) and 
the typical trajectories that a hard minor circular arc follows in the filamentary phase 
(Section \ref{nonema2}).
While in the lower density isotropic phase and in the (quasi-)nematic phase 
the positional one-particle density function is constant,
in the filamentary phase 
the positional one-particle density function is undulate
as in a smectic phase in three dimensions
(Section \ref{nonema1}).
Similar to 
the motion that a (hard) elongate particle makes 
across the layers in a smectic phase in three dimensions 
\cite{diffuSAesp,diffuSAsim}
is 
the motion that a hard minor circular arc makes 
across the filaments in a filamentary phase 
(Section \ref{nonema2}).

In their support of the non-nematicity of the filamentary phase in systems of hard minor circular arcs,
the results of the present article raise two doubts.
One, more general, is as to 
whether 
a ``modulated'' [``splay(-bend)''], truly nematic, phase could ever exist 
or, instead, 
whether, for that (polar, nematic) director periodicity to exist, 
it has to be necessarily associated to a local density periodicity.
The other, more particular, is as to 
whether 
a (lower-)virial density-functional theory,
which is capable to reproduce the filamentary phase,
could also be capable to reproduce 
the clustering in the high-density isotropic phase \cite{arcodue}
%that anticipates the formation of a filamentary phase,
or, instead,
whether that clustering is the symptom of the virial series expansion 
exhausting its convergency  
(Section \ref{conclusione}).

\section{
%the non-nematicity of the filamentary phase
results}
\label{nonema}
\subsection{one-particle density functions: second-virial density-functional theory and Monte Carlo numerical simulation}
\label{nonema1}

In the most general formulation 
of the second-virial density-functional theory, 
the free energy \(\cal{F}\) of a N--particle system is approximated as:
\begin{widetext}
\begin{equation}
\upbeta {\cal{F}}_{\rm IIvirial} = \int d{\mathtt x} \rho({\mathtt x}) \left[\log \left ({\cal V} \rho({\mathtt x}) \right) - 1 \right] +
\dfrac{1}{2} \int \int d{\mathtt x} d{\mathtt x}^{\prime} \rho({\mathtt x}) \rho({\mathtt x}^{\prime}) M({\mathtt x}, {\mathtt x}^{\prime})
\label{eq1} 
\end{equation}
\end{widetext}
in which:
\(\upbeta = 1/(k_B T)\), 
with \(k_B\) the Boltzmann constant and 
\(T\) the absolute thermodynamical temperature;
\(\mathtt x\) collects the positional, orientational and internal variables 
that contribute to define the mechanical state of a particle;
\(\rho ({\mathtt x})\) is the one-particle density function 
such that \[ \displaystyle \int d{\mathtt x} \rho({\mathtt x}) = {\rm N} \,;\]
\({\cal V}\) is the appropriate ``thermal'' (line, area, hyper-)volume; \( M({\mathtt x}, {\mathtt x}^{\prime}) \) is (\(-\)) the Mayer function such that
\begin{equation}
M({\mathtt x}, {\mathtt x}^{\prime}) = 1 - \mathsf{e}^{-\upbeta u ({\mathtt x}, {\mathtt x}^{\prime})},
\end{equation}
with \( u ({\mathtt x}, {\mathtt x}^{\prime}) \) the two-particle interaction potential energy function \cite{gomper}.

It is expected 
that  Eq. \ref{eq1} is a valid approximation to the exact \(\cal F\) 
only in the limit \(\uprho \rightarrow 0\):
only under these conditions can 
the higher-order terms 
of the virial series expansion of \(\cal F\) 
be safely neglected. 

So dilute, a system regularly is 
in the completely disordered isotropic phase.
There is an exception:
for a system of hard, long and thin, elongate particles in three dimensions,
as their aspect ratio diverges to infinity, 
Eq. \ref{eq1} becomes more and more capable to describe the exact \(\cal F\) also 
if it is 
in the positionally disordered but orientationally ordered nematic phase and 
thereby Eq. \ref{eq1} becomes capable to describe 
the (first-order) transition between 
these two positionally uniform phases,
which occurs at values of \(\uprho\) that converge to zero \cite{onsager}.
Instead, in a system 
of hard infinitesimally--thin disc-like particles in three dimensions,
the nematic phase becomes thermodynamically stable at finite \(\uprho\) and 
thereby the (first-order) isotropic--nematic phase transition occurs 
at finite \(\uprho\) \cite{eppengadisc} 
so that, under these conditions, Eq. \ref{eq1} is an invalid approximation to the exact \(\cal F\). 

Even in the two--dimensional analogue 
of a three--dimensional system 
of hard infinitesimally--thin disc-like particles, namely, 
a system of hard segments, 
i.e., hard circular arcs with \(\uptheta \rightarrow 0\), 
the (quasi-)nematic phase becomes (thermodynamically) stable at finite \(\uprho\) and 
thereby the isotropic--(quasi-)nematic phase transition occurs 
at finite \(\uprho\) \cite{eppengaseg}
so that, under these conditions, Eq. \ref{eq1} would, stricto sensu, be 
an invalid approximation to the exact \(\cal F\).

This notwithstanding, 
the comparison of the equation of state 
that the second-virial density-functional theory produces
for a system of hard segments, i.e., hard circular arcs with \(\uptheta \rightarrow 0\), 
to the corresponding equation of state 
from Monte Carlo numerical simulations 
is rather favorable \cite{polacca}.

With the confidence that such a favorable comparison provides but 
%with the clear proviso 
with the consciousness
that a second-virial density-functional theory can generally provide, 
at best, a coarse, ``impressionistic'', view on 
the thermodynamics of a system of hard particles,
Eq. \ref{eq1} is 
%cautiously 
used to investigate 
the phases and the transitions between them
in systems of hard minor circular arcs.

\begin{figure*}
\includegraphics[scale=0.75]{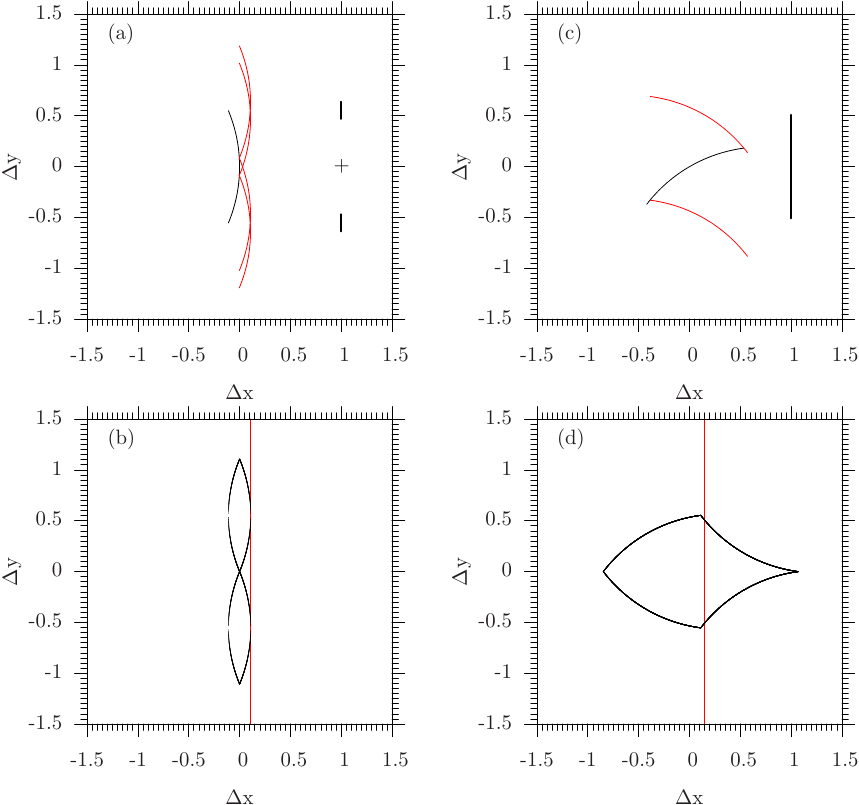}
\caption{
Illustration of the meaning of \({\cal S} (\Delta {\rm x}, \upvarphi, \upvarphi^{\prime})\).
In (a,b) \(\uptheta=\uppi/4\) and \(\upvarphi =\upvarphi^{\prime} = 0\);
in (a) there are, for example, the four limit pair configurations at which 
the two circular arcs 
(black, the one at the center, and red or gray, the one around) 
begin or cease to overlap: 
\({\cal S} (\Delta {\rm x}, \upvarphi, \upvarphi^{\prime})\) 
returns the sum of the lengths of the two segments that are shown on the right;
in (b), equivalently, take the excluded area (interior of the black curve) and 
cut it by a straight line parallel to the \textsl{y} axis (red or gray): 
the total length of the segments that constitute 
the interception of  this straight line with the excluded area provides 
\({\cal S} (\Delta {\rm x}, \upvarphi, \upvarphi^{\prime})\).
In (c,d) \(\uptheta=\uppi/4\) and \(\upvarphi = 2 \uppi/3 \) and \(\upvarphi^{\prime} = \uppi/3\);
in (c) there are, for example, the two limit pair configurations at which 
the two circular arcs (black, the one at the center, and red or gray, the one around) 
begin or cease to overlap: 
\({\cal S} (\Delta {\rm x}, \upvarphi, \upvarphi^{\prime})\)
returns the length of the segment that is shown on the right;
in (d), equivalently, take the excluded area (interior of the black curve) and 
cut it by a straight line parallel to the \textsl{y} axis (red or gray): 
the length of the segment that constitutes 
the interception of  this straight line with  the excluded area provides 
\({\cal S} (\Delta {\rm x}, \upvarphi, \upvarphi^{\prime})\).
}
\label{figura3}
\end{figure*}

In the filamentary phase of a system of hard minor circular arcs,
once the axes of the filaments have been taken along the \textsl{y} axis and 
it is, for reasonableness and simplicity, assumed that its periodicity exactly equals \(4R\), 
with \(R\) the radius of the parent circumference (Fig. \ref{figura2}),
Eq. \ref{eq1} becomes:
\begin{figure}[h!]
\includegraphics[scale=1]{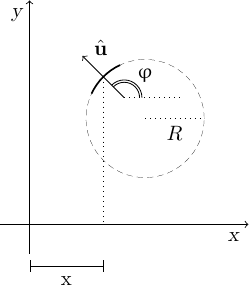}
\caption{
Example of a circular arc (continuous line)
in a (\textsl{x},\textsl{y}) plane with:
x the abscissa of its vertex;
\({\hat{\mathbf{u}}}\) the unit vector
which 
forms an angle \(\upvarphi\) with the \textsl{x} axis and
lies on the direction and sense
from the center of its parent circumference (discontinuous line),
whose radius is \(R\),
to its vertex.
%\(R\) is the radius of the parent circumference (discontinuous line).
}
\label{figura2}
\end{figure}
\begin{widetext}
\begin{eqnarray}
\upbeta {\mathsf{f}_{F}} = 
\log{\cal V} + \log \uprho - 1 + 
\dfrac{1}{4R}
\int_{-2R}^{+2R} d {\rm x} \, \int_0^{2\uppi} d\upvarphi
G({\rm x},\upvarphi) \log G({\rm x}, \upvarphi) + \nonumber \\
+\frac{1}{2} \, \uprho \,
\dfrac{1}{4R}
\int_{-2R}^{+2R} d{\rm x}\, 
\int_0^{2\uppi} d\upvarphi 
G({\rm x},\upvarphi) 
\int 
d{\Delta \rm x} \int_0^{2\uppi} d\upvarphi^{\prime}
G({\rm x}+\Delta{\rm x},\upvarphi^{\prime}) 
{\cal S} (\Delta {\rm x}, \upvarphi, \upvarphi^{\prime})
\label{eq2}
\end{eqnarray}
\end{widetext}
in which:
\({\mathsf{f}} = {\cal F}/{\rm N}\);
\({\rm x} \) is the abscissa of the vertex of a circular arc (Fig. \ref{figura2});
\(\upvarphi\) is the angle
that the unit vector \(\hat {\mathbf u}\)
forms with the \textsl{x} axis, 
\(\hat {\mathbf u} \) lying on the direction and sense  
from the center of the parent circumference of radius \(R\)  
to the vertex of the circular arc (Fig. \ref{figura2});
\(G({\rm x}, \upvarphi)\) is the probability density function to find
a hard minor circular arc 
whose vertex has abscissa \({\rm x}\) and orientation angle is \(\upvarphi\);
%and is normalized such that
%\[
%\frac{1}{4R} \int_{-2R}^{+2R} d{\rm x} \int_0^{2\uppi} d\upvarphi
%G({\rm x}, \upvarphi) = 1\,;
%\]  
%\(d_{\rm max}\) is the distance along the x axis
%between two circular arc beyond which it is certain
%that they do not overlap;
\begin{equation}
{\cal S} (\Delta {\rm x}, \upvarphi, \upvarphi^{\prime}) = 
\int d{\Delta {\rm y}} M({\Delta {\rm x}}, {\Delta {\rm y}}, 
\upvarphi, \upvarphi^{\prime})
\label{eq3}
\end{equation} 
which is the  (total) length of the segment(s) 
that the vertex of the circular arc with 
position \(({\Delta {\rm x}}, {\Delta {\rm y}})\) and
orientation \(\upvarphi^{\prime}\) 
spans 
while it overlaps with a circular arc
with position \((0,0)\) and orientation \(\upvarphi\); i.e.,
equivalently,
the (total) length of the segment(s) of the interception 
of the excluded area 
between two hard circular arcs with orientations \(\upvarphi\) 
and \(\upvarphi^{\prime}\) with a straight line 
that is displaced by \(\Delta {\rm x}\) from 
the vertex of the hard circular arc with orientation \(\upvarphi\)
and is 
parallel to the \textsl{y} axis (Fig. \ref{figura3}).

If \(G({\rm x},\upvarphi) = f(\upvarphi)\), 
%\(f(\upvarphi)\) being the orientational probability density function,
%and the integrals over \(\rm x\) and \(\Delta {\rm x}\) are evaluated,
Eq. \ref{eq2} reduces to the free energy per particle 
of the nematic phase in the second-virial approximation:
\begin{widetext}
\begin{equation}
\upbeta \mathsf{f}_N = \log {\cal V} + \log \uprho - 1 + \int_0^{2\uppi} f(\upvarphi) \log f(\upvarphi) + 
\frac{1}{2}  \uprho \int_0^{2\uppi} d\upvarphi \int_0^{2\uppi} d\upvarphi^{\prime} f(\upvarphi) f(\upvarphi) \, a(\upvarphi, \upvarphi^{\prime})
\label{eq4}
\end{equation}
\end{widetext}
with \(a(\upvarphi, \upvarphi^{\prime}) \) the excluded area between two hard circular arcs with 
orientations \(\upvarphi\) and \(\upvarphi^{\prime}\) [Fig. \ref{figura3} (b,d)].

If \(f(\upvarphi) = 1/(2\uppi)\), Eq. \ref{eq4} reduces to the free energy per particle of
the isotropic phase in the second-virial approximation:
%\begin{widetext}
\begin{equation}
\upbeta \mathsf{f}_I = \log {\cal V} + \log \uprho - 1 -\log (2\uppi) + 
\frac{1}{2} \uprho \langle  a \rangle 
\label{eq5}
\end{equation}
%\end{widetext}
with \(\langle  a \rangle\) the completely orientationally averaged excluded area.

\begin{figure}[h!]
\includegraphics[scale=0.75]{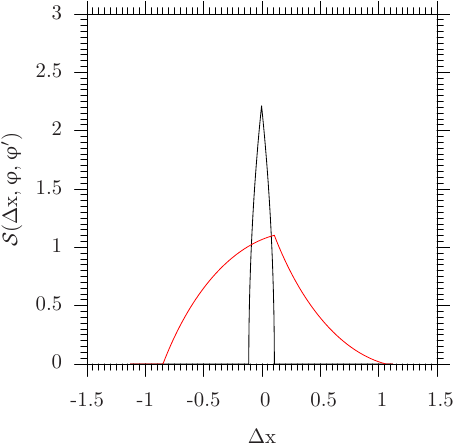}
\caption{
Examples of the graph of the function 
\({\cal S} (\Delta {\rm x}, \upvarphi, \upvarphi^{\prime})\).
Specifically, \(\cal S\) as a function of \(\Delta {\rm x}\) 
with \(\upvarphi = \upvarphi^{\prime} = 0 \) (black) 
and 
\(\upvarphi = 2\uppi/3\) and \(\upvarphi^{\prime} = \uppi/3\) (red or gray).
}
\label{figura4}
\end{figure}

For any value of  \(\uptheta\) that is considered, 
one has to determine the minimum 
of \(\mathsf{f}_F\) and of \(\mathsf{f}_N\) as well as \(\mathsf{f}_I\) 
for as many values of \(\uprho\) as necessary.
To this aim, 
one has to numerically construct the function 
\({\cal S} (\Delta {\rm x}, \upvarphi, \upvarphi^{\prime})\) 
in Eq. \ref{eq3} (Fig. \ref{figura4}), 
which constitutes the ``kernel'' of Eq. \ref{eq2}.
From  \({\cal S} (\Delta {\rm x}, \upvarphi, \upvarphi^{\prime})\),
one determines \(a(\upvarphi, \upvarphi^{\prime})\) as
\begin{equation}
a(\upvarphi, \upvarphi^{\prime}) = \int d\Delta {\rm x} {\cal S} (\Delta {\rm x}, \upvarphi, \upvarphi^{\prime})
\label{eq6}
\end{equation}
(Fig. \ref{figura3}), which is the ``kernel'' of Eq. \ref{eq4}.
From \(a(\upvarphi, \upvarphi^{\prime})\), 
one determines \(\langle a \rangle\)  as
\begin{equation}
\langle a \rangle = \frac{1}{(2\uppi)^2} \int_0^{2\uppi} d\upvarphi \int_0^{2\uppi} d\upvarphi^{\prime} a(\upvarphi, \upvarphi^{\prime})
\label{eq7}
\end{equation}
which is the key quantity that enters Eq. \ref{eq5}. 

Once \( \langle a \rangle \) is available, 
\(\mathsf{f}_I\) is determined by  
directly evaluating Eq. \ref{eq5} for 
as many values of \(\uprho\) as necessary. 
Pressure \(P_I (\uprho) \) and chemical potential \(\upmu_I (\uprho) \) of the isotropic phase can be determined
by suitable thermodynamical differentiation of Eq. \ref{eq5}.

The determination 
of the minimum of \(\mathsf{f}_N\), which functionally depends on \(f(\upvarphi)\), and 
of the minimum of \(\mathsf{f}_F\), which functionally depends on \(G({\rm x}, \upvarphi)\), 
amounts to constrained-minimizing Eq. \ref{eq4} and Eq. \ref{eq2} 
with respect to, respectively,  \(f(\upvarphi)\) and \(G({\rm x}, \upvarphi)\).
The mathematical rigorous way to achieve these constrained minimizations conduces to 
the two respective non-linear integral equations:
\begin{equation}
\log \left [ J f(\upvarphi) \right] = - \uprho \int_0^{2\uppi} d\upvarphi^{\prime} f(\upvarphi^{\prime})
a(\upvarphi, \upvarphi^{\prime}) 
\label{eq8}
\end{equation}
with \(J\) a constant that is determined by the normalization condition 
\[\int_0^{2\uppi} f(\upvarphi) = 1 \,;\]
\begin{widetext}
\begin{equation}
\log \left [ K G({\rm x}, \upvarphi) \right] = - \uprho  \int d \Delta {\rm x} \int_0^{2\uppi} d\upvarphi^{\prime} 
G({\rm x}+\Delta{\rm x}, \upvarphi^{\prime}) {\cal S} (\Delta {\rm x}, \upvarphi, \upvarphi^{\prime})
\label{eq9}
\end{equation}
\end{widetext}
with \(K\) a constant that is determined by the normalization condition 
\[
\frac{1}{4R} \int_{-2R}^{+2R} d{\rm x} \int_0^{2\uppi} d\upvarphi
G({\rm x}, \upvarphi) = 1\,.
\] 

The non-linearity 
of the two integral equations in Eq. \ref{eq8} and Eq. \ref{eq9} 
forces one to search for their solution
by implementing a suitable iterative method 
similar to the one that was implemented for the solution 
of the ``Onsager'' integral equation in three dimensions 
\cite{herzfeld}.

Similarly to this past calculation,
the iterative method to solve Eq. \ref{eq8} was relatively easy.
For any value of \(\uptheta\) that was considered, 
one could commence with the initial probability density function
\[
f_0 \left(\upvarphi
%=\cos^{-1}( \hat{\mathbf u}\cdot\hat{n}) 
\right) \propto 
\mathsf{e}^{\uplambda\,(2(\hat{\mathbf u}(\upvarphi) \,\cdot\, \hat{n})^2 - 1)} \;,
\]
with, e.g., \(\hat{n}=(0,\pm 1)\) and \(\uplambda = 10\),
at a sufficiently large value of \(\uprho\).
The probability density function at iteration \(k+1\) was so related to 
the probability density function at iteration \(k\):
\[
f_{k+1} (\upvarphi) = 
\dfrac{\mathsf{e}^{-\uprho \, \int_0^{2\uppi} d \upvarphi^{\prime} f_k(\upvarphi^{\prime}) a(\upvarphi,\upvarphi^{\prime})}}
{\int_0^{2\uppi} d\upvarphi \,
\mathsf{e}^{-\uprho \, \int_0^{2\uppi} d \upvarphi^{\prime} f_k(\upvarphi^{\prime}) a(\upvarphi,\upvarphi^{\prime})}} \;,
\]
with the relevant integrals 
that were evaluated 
by the simple mid-point rectangle method.
The iterations ceased as soon as the assumed convergence criterion 
\[
\left | \max\limits_{\upvarphi \in [0,2\uppi]} f_{k+1} (\upvarphi) - \max\limits_{\upvarphi \in [0,2\uppi]} f_{k} (\upvarphi) \right| < 10^{-7} 
\] 
was satisfied. 
Usually, the convergence was very rapid. 
The final probability density function 
at a certain value of \(\uprho\) 
was taken as
the initial probability density function 
at the immediately smaller value of \(\uprho\) 
that was considered. 
In this way, the probability density function \(f(\upvarphi)\) 
at many values of \(\uprho\)
%from the higher-density nematic phase to the lower-density isotropic phase,
was determined.
From Eq. \ref{eq4}, \(\mathsf{f}_N\) at these values of \(\uprho\) was 
evaluated.  
Pressure \(P_N (\uprho) \) and chemical potential \(\upmu_N (\uprho) \) of the nematic phase were determined 
by suitable thermodynamical differentiation of Eq. \ref{eq4}.
For any value of \(\uptheta\) that was considered,
there exists a special value of \(\uprho\), \(\uprho_{IN}\), such that,
for \(\uprho < \uprho_{IN}\), only the isotropic solution \(f_I (\upvarphi) = 1/(2\uppi)\) to Eq.  \ref{eq8}  exists.
The special value \(\uprho_{IN}\) is the value of \(\uprho\)  at which
the nematic solution \(f_N (\upvarphi)\) ``bifurcates'' off the isotropic solution \(f_I (\upvarphi)\);
it can be analytically determined by bifurcation analysis \cite{rave}:
\begin{equation}
\uprho_{IN} = - \frac{2\uppi}{\int_0^{2\uppi} d \upgamma a (\upgamma) \cos (2\upgamma)}
\end{equation}  
with \(\upgamma = \upvarphi^{\prime} - \upvarphi\),
as \(a\) actually depends on the angle comprised
between the orientation angles, \(\upvarphi\) and
\(\upvarphi^{\prime}\), of the two hard circular arcs.
In common with a system of hard segments, i.e., hard circular arcs with \(\uptheta \rightarrow 0\) \cite{rave},
the second-virial density functional theory finds that 
a second-order (continuous) isotropic-nematic phase transition  occurs at \(\uprho_{IN}\).
Incidentally, it is pertinent to note that the vicinity of the isotropic-nematic phase transition was revealed by 
the increase in the number of iterations that were necessary to achieve convergence
as \(\uprho\) approached \(\uprho_{IN}\).

The iterative method to solve Eq. \ref{eq9} was less easy. 
For any value of \(\uptheta\) and a certain value of \(\uprho\), 
the initial probability density function was either 
\[
G_0 ({\rm x}, \upvarphi) \propto 
\mathsf{e}^{\uplambda \hat{\mathbf u}(\upvarphi) \,\cdot\, \hat{\mathbf p }({\rm x})} \,, 
\]
with
\begin{widetext}
\begin{eqnarray*}
{\hat{\mathbf p}} ({\rm x}) = (p_x, p_y)=
\begin{cases}
p_{x} = \dfrac{{\rm x} -2jR}{R} \,, (2j-1)R \leq {\rm x} \leq (2j+1)R & \\
& ,\, \forall j\in\mathbb{Z}\,, \\
p_{y} = (-1)^{j} \sqrt{1-p_{x}^2}  & 
\end{cases}
\end{eqnarray*}
\begin{figure}[h!]
\includegraphics[scale=1]{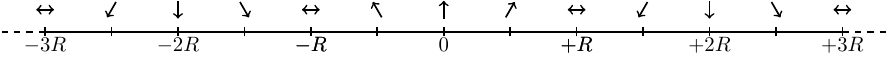}
\caption{Schematic 
of the variation of \(\hat{\mathbf p} \) along the axis \textsl{x} 
in a prototypical filamentary phase. 
Note the double-headed arrow 
at \({\rm x} = (2j+1)R\), \(j\in\mathbb{Z}\), 
implying local non-polarity.}
\label{figura5}
\end{figure}
\end{widetext}
to emulate the polar nematic director profile of a filamentary phase 
[cf. Fig. \ref{figura1}(c) with Fig. \ref{figura5}],
or a probability density function to which 
the iterations had converged at a close-by value of \(\uprho\).
It proved necessary to take 
the probability density function at iteration \(k+1\) 
as a mixture of 
the probability density function at iteration \(k\) and 
the provisional probability density function at iteration \(k+1\) 
\begin{widetext}
\[
{\mathsf{G}}_{\mathsf{k+1}} ({\rm x}, \upvarphi) = 
\dfrac{ 
\mathsf{e}^{-\uprho\, \int d\Delta {\rm x} \int_0^{2\uppi} d\upvarphi^{\prime} 
G_k ({\rm x} + \Delta{\rm x}, \upvarphi^{\prime} ) 
{\cal S} (\Delta {\rm x}, \upvarphi, \upvarphi^{\prime})}}
{\int d{\rm x} \int_0^{2\uppi} d\upvarphi \,
\mathsf{e}^{-\uprho\, \int d\Delta {\rm x} \int_0^{2\uppi} d\upvarphi^{\prime} 
G_k ({\rm x} + \Delta{\rm x}, \upvarphi^{\prime} )  {\cal S} (\Delta {\rm x}, \upvarphi, \upvarphi^{\prime})}}
 \,,
\]
\end{widetext}
with the relevant integrals 
that were evaluated 
by the simple mid-point rectangle method.
Specifically, it was taken 
\[
G_{k+1} ({\rm x}, \upvarphi) = \frac{1}{4} {\mathsf{G}}_{\mathsf{k+1}} ({\rm x}, \upvarphi) + \frac{3}{4} G_{k} ({\rm x}, \upvarphi)
\]
to ensure convergence which, albeit relatively slow, was 
to a physically acceptable probability density function
compatible with the structure of a filamentary phase.
The weight of \( {\mathsf{G}}_{\mathsf{k+1}} ({\rm x}, \upvarphi) \) 
equal to \(1/4\), and hence that of \(G_{k} ({\rm x}, \upvarphi)\) equal to \(3/4\), 
was empirically found as
a reasonable compromise between 
the physicality of the solution and the rapidity of the convergence.
If the weight of 
\( {\mathsf{G}}_{\mathsf{k+1}} ({\rm x}, \upvarphi) \) was much larger, 
e.g. equal to 1, and hence that of \(G_{k} ({\rm x}, \upvarphi)\) equal to \(0\), 
the convergence was significantly more rapid but 
it was to a physically unacceptable probability density function
compatible with the structure of a layered phase with isotropic layers.
%The smaller was the weight of \(G_{``k+1''} ({\rm x}, \upvarphi)\), 
%the slower was the convergence which, however, was to 
%a physically acceptable probability density function compatible with the structure of a filamentary phase.
The iterations ceased as soon as the assumed convergence criterion
\begin{widetext} 
\[
\left | 
\max\limits_{
\begin{cases}
{\rm x} \in [-2R,+2R] \\
\upvarphi \in [0,2\uppi]
\end{cases}
} 
G_{k+1} ({\rm x}, \upvarphi) - 
\max\limits_{
\begin{cases}
{\rm x} \in [-2R,+2R] \\
\upvarphi \in [0,2\uppi]
\end{cases}
} 
G_{k} ({\rm x}, \upvarphi) 
\right| < 10^{-7} 
\] 
\end{widetext}
was satisfied.
In this way, the probability density function \(G  ({\rm x}, \upvarphi) \) 
at many values of \(\uprho\) 
was determined.
From Eq. \ref{eq2}, \(\mathsf{f}_F\) 
at these values of \(\uprho\) 
was evaluated.
Pressure \(P_F (\uprho) \) and chemical potential \(\upmu_F (\uprho)\) 
were determined
by suitable thermodynamical differentiation of Eq. \ref{eq2}. 
It is pertinent to note that,
at sufficiently high density, 
the values of \(P_F\) and \(\upmu_F\) 
that resulted from the physical solution 
were lower than 
the values of \(P_N\) and \(\upmu_N\) 
at the same value of \(\uprho\);
instead, the values of pressure and chemical potential 
that resulted from the unphysical solution 
were significantly larger than 
the values of \(P_N\) and \(\upmu_N\) 
at the same value of \(\uprho\). 
Once a sufficiently small value of \(\uprho\) was considered,
the iterations converged to the nematic solution 
\(G({\rm x}, \upvarphi) = f_N(\upvarphi)\).
By reducing \(\uprho\) further, 
the iterations finally converged to the isotropic solution 
\(G({\rm x}, \upvarphi) = f_I(\upvarphi) = 1/(2\uppi)\).
Incidentally, it is pertinent to note
that such changes of the nature of the solution were in correspondence
of values of \(\uprho\) at which convergence was particularly slow.

Once the three possible branches, 
isotropic, nematic and filamentary, 
of the free energy 
had been determined,
the one branch that had the minimal \(\mathsf{f}\) was selected 
as the thermodynamically stable phase at a certain value of \(\uprho\) 
as well as  
transitions between two phases were determined 
by searching for 
what values of \(\uprho\) 
there was equality of pressure and chemical potential.
In this way, 
the complete ``second-virial density-functional theory phase diagram'' 
in the \((\uptheta,\uprho)\) plane
could have been traced.
This task might be completed then;
now, the focus is 
on those particular values of \(\uptheta\),
\(\uptheta = 0.5\), \(\uptheta = 1\) and \(\uptheta =1.8\),
for which Monte Carlo numerical simulation results are available \cite{arcodue}
%that reveal the formation of a filamentary phase at higher density 
%and thus 
that allow for
a direct comparison with results 
from second-virial density-functional theory. 
%is possible.

\begin{figure}[h!]
\includegraphics{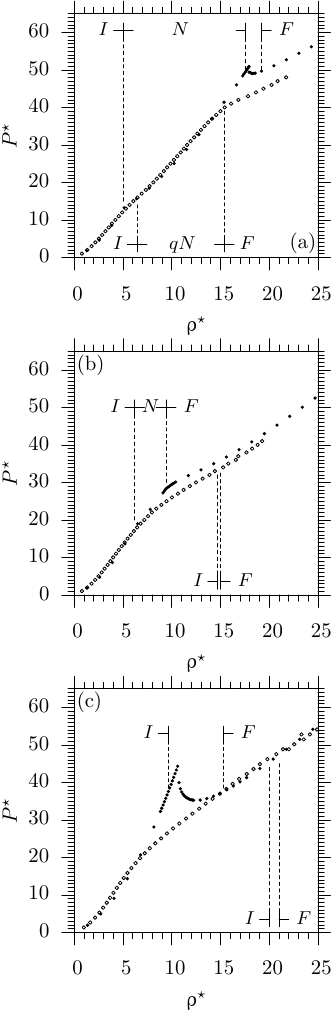}
\caption{Equation of state, dimensionless pressure 
\(P^{\star} = \upbeta P \ell^2\) versus 
dimensionless number density \(\uprho^{\star} 
=\uprho \ell^2\), with \(\ell\) the length 
of a circular arc,  for a system of
hard circular arcs with \(\uptheta=0.5\) (a), 
\(\uptheta = 1\) (b) and \(\uptheta = 1.8\) (c),
from second-virial density-functional theory (black circles) and 
Monte Carlo numerical simulations (white circles). 
In each panel, on the top there is 
the sequence of phases 
that the second-virial density-functional theory has obtained 
while on the bottom there is 
the sequence of phases 
that the Monte Carlo numerical simulations obtained \cite{arcodue}, 
with \(I\) the isotropic phase, 
\((q)N\) the (quasi-)nematic phase and 
\(F\) the filamentary phase.}
\label{figura6}
\end{figure}
Similarly to what occurs for a system of hard segments, i.e., 
hard circular arcs with \(\uptheta \rightarrow 0\) \cite{polacca},
the sequence of the phases and their equation of state
that the second-virial density functional theory produces
for a system of hard circular arcs with \(\uptheta = 0.5\)
accord reasonably well with
the sequence of phases and their equation of state
that were observed in 
the Monte Carlo numerical simulations 
for the same system of hard minor circular arcs [Fig. \ref{figura6} (a)] \cite{arcodue}.
(Leave aside the issue 
of the location 
of the isotropic--(quasi-)nematic phase transition
and
the issue of 
the nature of the (quasi-)nematic phase in two dimensions:
the second-virial density functional theory
can only deal with an ordinary nematic phase
that becomes stabler than the isotropic phase 
beyond a second-order phase transition at \(\uprho_{IN}\);
in the Monte Carlo numerical simulations
the criterion for 
the transition from
the isotropic phase to the (quasi-)nematic phase 
was pragmatically based on
the long-distance decay of
the second-order orientational
two-particle correlation function \cite{arcodue}.) 
The principal difference between 
the second-virial density functional theory results and
the Monte Carlo numerical simulation results
concerns the transition between the (quasi-)nematic
phase and the filamentary phase:
the Monte Carlo numerical simulations found it 
second-order (continuous) \cite{arcodue}, 
whereas
the second-virial density-functional theory finds it 
first-order (discontinuous) 
and
at values of \(\uprho\) larger than the value of \(\uprho\) at which 
the (quasi-)nematic phase transits to the filamentary phase 
in the Monte Carlo numerical simulations \cite{arcodue}.

The accord between 
the second-virial density-functional theory and 
the Monte Carlo numerical simulations 
deteriorates for a system of hard circular arcs with \(\uptheta=1\) 
[Fig. \ref{figura6} (b)]. 
Not only because the two respective equations of state differ but also, and especially, 
because the second-virial density-functional theory continues to find 
a nematic phase in between 
the isotropic phase and the filamentary phase, 
with both isotropic--nematic and nematic--filamentary phase transitions 
being second-order (continuous), 
whereas the Monte Carlo numerical simulations
found a weakly first-order (discontinuous) 
isotropic--filamentary phase transition  
at values of \(\uprho\) larger than 
the value of \(\uprho\) at which the filamentary phase
begins to be thermodynamically stable 
in the second-virial density-functional theory \cite{arcodue}. 
Much of the responsibility for this discrepancy should rest on 
the intrinsic incapability 
of a (second-)virial  density-functional theory 
to deal with the no ordinary cluster isotropic phase
that the  Monte Carlo numerical simulations revealed 
at higher density \cite{arcodue}.

Second-virial density-functional theory and Monte Carlo numerical simulations 
return to being more accordant
for a system of hard circular arcs with \(\uptheta=1.8\) [Fig. \ref{figura6} (c)]. 
Even though the two equations of state significantly differ,
at least the same sequence of phases is found by both methods:
an isotropic phase at lower density is followed by 
a filamentary phase at higher density, 
the two phases being separated 
by a first-order (discontinuous) phase transition;
the latter is, expectedly, much stronger and occurs at smaller values
of \(\uprho\)
in the second-virial density-functional theory than 
in the Monte Carlo numerical simulations
for the intrinsic incapability of the former to deal with 
the no ordinary cluster isotropic phase that 
the latter revealed at higher density \cite{arcodue}.  

Irrespective of the (in)favorability 
of the quantitative comparison between 
the second-virial density-functional theory results and 
the Monte Carlo numerical simulation results, 
it is qualitatively relevant that 
also the second-virial density-functional theory is capable 
to find a filamentary phase at higher density. 

\begin{figure}[h!]
\includegraphics[scale=0.9]{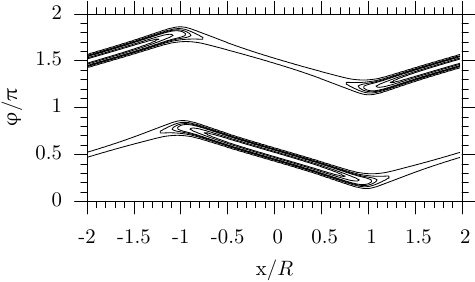}
\caption{
Example of 
the contour plot of 
\(G({\rm x}, \upvarphi)\) in the filamentary phase 
of a system of hard minor circular arcs 
from second-virial density-functional theory.
Specifically, 
the filamentary phase 
of a system of hard circular arcs with \(\uptheta = 0.5\) 
at \(\uprho^{\star}   = \uprho \ell^2 = 20.5\), 
with \(\ell \) the length of a circular arc;
the contours are drawn for 
the values of \(G({\rm x}, \upvarphi) \) 
equal to 0.25, 0.5, 0.75, 1, 2 and 3.
} 
\label{figura7}
\end{figure}

\begin{figure}[h!]
\includegraphics{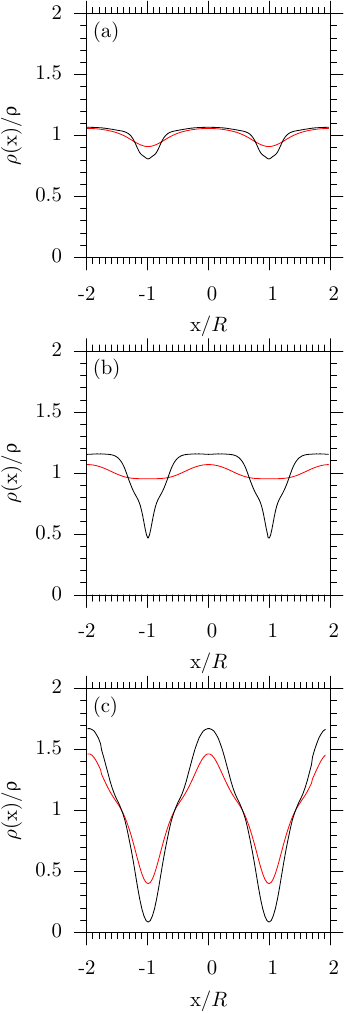}
\caption{
Examples of the graph of \(\rho({\rm x})/\uprho\) in
the filamentary phase of a system of hard minor circular arcs 
from second-virial density-functional theory at values
of \(\uprho^{\star} = \uprho \ell^2\), with \(\ell\) 
the length of a circular arc,
close to, and far from, the transition to the isotropic phase or the nematic phase.
Specifically, a system of hard circular arcs with:
(a) \(\uptheta=0.5\) at values
of \(\uprho^{\star} = 19.2\) 
(red or gray) and 
\(\uprho^{\star} =  24.3\) (black);
(b) \(\uptheta=1\) at values 
of \(\uprho^{\star} =  10.4\) (red or gray) and
\(\uprho^{\star} =  24.7\) (black); 
(c) \(\uptheta=1.8\) at values
of \(\uprho^{\star} =  15\) (red or gray) and
\(\uprho^{\star} =  24.5\) (black).
}
\label{figura8}
\end{figure}

Similarly to any density-functional theory \cite{wu}, 
the principal result 
of the present second-virial density-functional theory
is the corresponding one-particle density function:
\begin{equation}
\rho({\rm x}, \upvarphi) / \uprho = G({\rm x}, \upvarphi)  \,.
\end{equation}
In the filamentary phase, 
the contour plot of \(G({\rm x}, \upvarphi)\) could superficially be interpreted
as consistent with a ``modulated'' [``splay(-bend)''] nematic phase:
the values of \(\upvarphi\) 
at which  \(G({\rm x}, \upvarphi)\) is locally maximal 
undulately vary with \(\rm x\) (Fig. \ref{figura7}).
For the filamentary phase to qualify for being nematic,
these undulations must be compatible with 
a constant positional one-particle density function:
\begin{equation}
\rho({\rm x})/\uprho = \int_0^{2\uppi} d\upvarphi  G({\rm x}, \upvarphi).
\end{equation}
Instead, for any value of \(\uptheta\), 
\(\rho({\rm x})/\uprho\) in the filamentary phase is always undulate, 
with period naturally equal to \(2R\), 
as in a smectic phase in three dimensions (Fig. \ref{figura8}).

\begin{figure}[h!]
\includegraphics[scale=0.9]{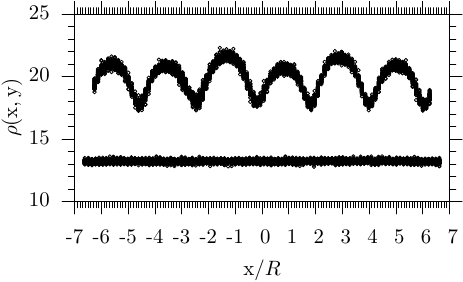}
\caption{
Local number density \(\rho({\rm x},{\rm y})\) (white circles)
as obtained 
in canonical Monte Carlo numerical simulations 
of a system of \({\rm N}=5400\) hard circular arcs with \(\uptheta=0.5\) 
in the (quasi-)nematic phase
at %\(P^{\star} = \upbeta P \ell^2 = 35\) (bottom) and 
\(\uprho^{\star} = \uprho \ell^2 = 13.21 \) (bottom) and
in the filamentary phase at
%\(P^{\star} = \upbeta P \ell^2 = 45\) (top), 
\(\uprho^{\star} = \uprho \ell^2 =19.86 \) (top),
with \(\ell\) the length of a circular arc,
as projected on the \(({\rm x},\rho)\) plane so as
to illustrate the uniformity of the former phase
and the non-uniformity along the axis transverse
to the axes of the filaments of the latter phase.
}
\label{figura9}
\end{figure}

%By calculating suitable two-particle correlation functions, 
%the non-uniformity of the filamentary phase was already appreciated in
%the previous isobaric(-isothermal) Monte Carlo numerical simulations \cite{arcodue}.
%In the present canonical Monte Carlo numerical simulations, 
%that have been conducted according to the same computational protocol
%of the previous isobaric(-isothermal) Monte Carlo numerical simulations \cite{arcodue},
%the non-uniformity of the filamentary phase is further indicated 
%by the form of the local number density \(\rho({\rm x}, {\rm y})\):
%along the axis \textsl{x} transverse to the axis \textsl{y} of the filaments 
%a ``density wave'' is appreciated (Fig. \ref{figura9}).
Previously, the non-uniformity of the filamentary phase was already appreciated
by calculating suitable two-particle correlation functions in
isobaric(-isothermal) Monte Carlo numerical simulations \cite{arcodue}.
Presently, additional canonical Monte Carlo numerical simulations
were conducted for a system of hard circular arcs with \(\uptheta = 0.5\) 
at five values of \(\uprho\): 
\(\uprho^{\star} = \uprho \ell^2 =\) 
\(13.21 \);
\(19.86 \); 
\(25.47 \);
\(30.37 \);
\(36.91 \);
with \(\ell\) the length of a circular arc.
The configurations 
with which the present five Monte Carlo numerical simulations in the canonical ensemble  were initiated
were obtained in 
the previous numerical simulations 
that were conducted by 
the Monte Carlo method in the isobaric(-isothermal) statistical ensemble  
at the respective values of pressure:
\(P^{\star} =\upbeta  P \ell^2 =\)
\(35 \);
\(45 \); 
\(55 \);
\(60 \);
\(80 \) \cite{arcodue}.
The present five Monte Carlo numerical simulations were conducted
by the same computational protocol of the previous Monte Carlo numerical simulations \cite{arcodue}
except for the absence of any attempt to modify the container.
Presently,
the non-uniformity of the filamentary phase is further indicated 
by the form of the local number density \(\rho({\rm x}, {\rm y})\):
along the axis \textsl{x} transverse to the axis \textsl{y} of the filaments 
a ``density wave'' is appreciated (Fig. \ref{figura9}).

Thus, 
both second-virial density-functional theory and 
Monte Carlo numerical simulation
agree on 
the non-nematicity of the filamentary phase.
(In retrospect, the non-nematicity of the filamentary phase cannot surprise, 
being the filamentary phase of hard minor circular arcs 
in the two--dimensional Euclidean space \({\mathbb R}^2\) 
the analogue of the cluster columnar phase of hard spherical caps 
in the three--dimensional Euclidean space \({\mathbb R}^3\) 
\cite{notuccia}.)

This notwithstanding, 
it remains relevant to digress on 
what is the actual profile 
that \(\hat {\mathbf p}\) and \(\hat n\) adopt along the \textsl{x} axis and 
what are the values 
of the polar order parameter \(S_1({\rm x })\) and 
of the nematic order parameter \(S_2({\rm x })\) 
in the filamentary phase
that the second-virial density-functional theory produces.
 
\begin{figure}[h!]
\includegraphics{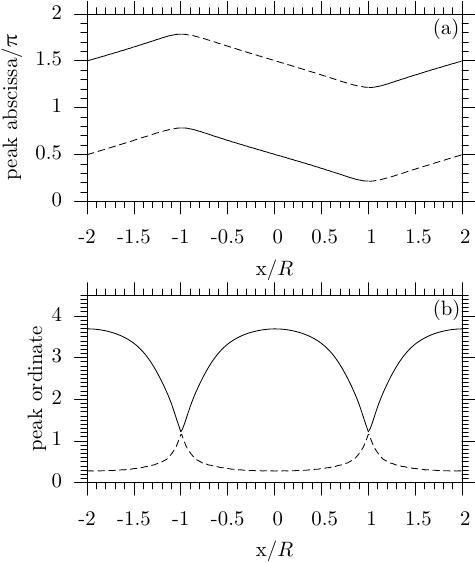}
\caption{
From the second-virial density-functional theory
for the filamentary phase 
of a system 
of hard circular arcs with \(\uptheta = 0.5\) 
at \(\uprho^{\star}   = \uprho \ell^2 = 20.5\), 
with \(\ell \) the length of a circular arc:
(a) the value of the abscissa  
at which \(f \left( \upvarphi \left| {\rm x} \right. \right)\)
has the major peak (continuous line) and the minor peak
(discontinuous line) as a function of \(\rm x\);
(b) the value of the ordinate of
the major peak 
(continuous line) and
of the minor peak (discontinuous line) of 
\(f \left( \upvarphi \left| {\rm x} \right. \right)\)
as a function of \(\rm x\). 
}
\label{figura10}
\end{figure}

\begin{figure}
\includegraphics{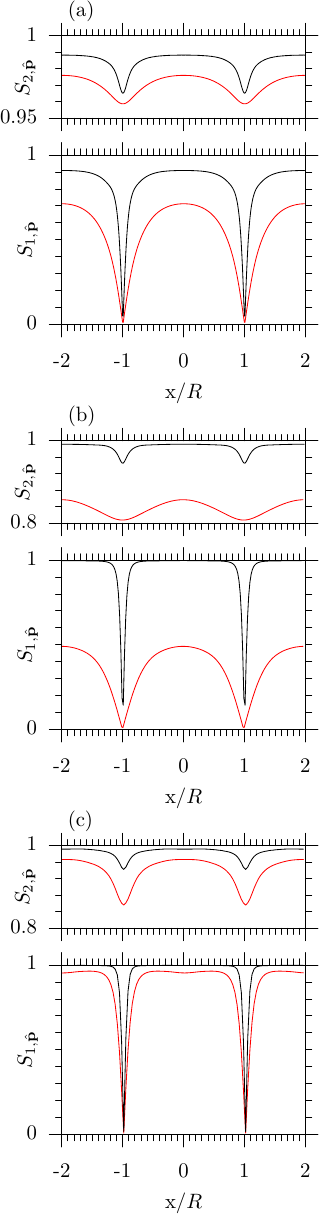}
\caption{
Examples of the graph of
\(S_{1,{\hat{\mathbf p}}}({\rm x})\) 
(bottom subpanel) and \(S_{2,{\hat{\mathbf p}}}({\rm x})\)
(top subpanel)
in
the filamentary phase of a system of hard minor circular arcs 
from second-virial density-functional theory at values
of \(\uprho^{\star} = \uprho \ell^2\), with \(\ell\) 
the length of a circular arc,
close to, and far from, the transition to the isotropic phase or the nematic phase.
Specifically, a system of hard circular arcs with:
(a) \(\uptheta=0.5\) at values
of \(\uprho^{\star} = 19.2\) 
(red or gray) and 
\(\uprho^{\star} =  24.3\) (black);
(b) \(\uptheta=1\) at values 
of \(\uprho^{\star} =  10.4\) (red or gray) and
\(\uprho^{\star} =  24.7\) (black); 
(c) \(\uptheta=1.8\) at values
of \(\uprho^{\star} =  15\) (red or gray) and
\(\uprho^{\star} =  24.5\) (black).
}
\label{figura11}
\end{figure}

\begin{figure}
\includegraphics{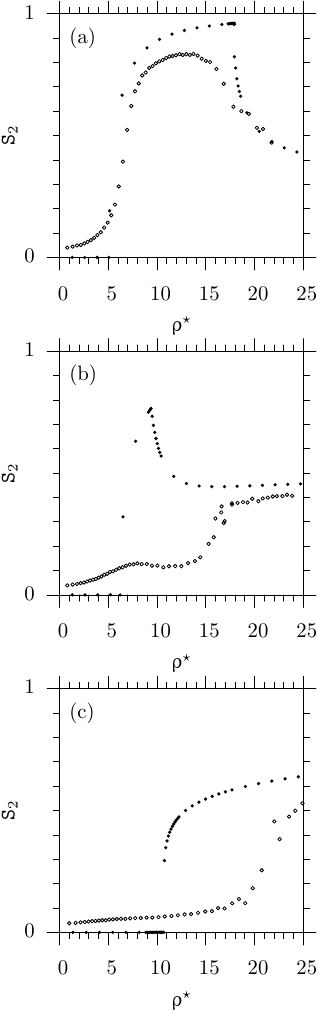}
\caption{
The nematic order parameter \({\mathtt{S}}_2 = 
\left \langle S_{2,{\rm {y}}}  \right \rangle \)
as a function of 
dimensionless number density \(\uprho^{\star} 
=\uprho \ell^2\), with \(\ell\) the length 
of a circular arc,  for a system of
hard circular arcs with \(\uptheta=0.5\) (a), 
\(\uptheta = 1\) (b) and \(\uptheta = 1.8\) (c),
from second-virial density-functional theory (black circles) and 
Monte Carlo numerical simulations \cite{arcodue} (white circles). 
}
\label{figura13}
\end{figure}

To this aim, 
one has to mathematically analyze 
the orientational distribution function as a function of \(\rm x\):
\begin{equation}
f \left( \upvarphi \left| {\rm x} \right. \right) = 
\dfrac{G({\rm x}, \upvarphi)}
{\int_0^{2\uppi} d\upvarphi G({\rm x}, \upvarphi)} \,.
\end{equation}
Irrespective of the value of \(\rm x\), 
\(f \left( \upvarphi \left| {\rm x} \right. \right)\) is bimodal
with a major peak and a minor peak (Fig. \ref{figura10}).
The abscissae and the ordinates of the two peaks concertedly vary along the \textsl{x} axis 
(Fig. \ref{figura10}).
The abscissa of the major peak defines \(\Uptheta\), 
the angle that \(\hat{\mathbf p}\) forms with the \textsl{x} axis.
In a filament, \(\Uptheta ({\rm x})\) is essentially linear with \(\rm x\) and,
on passing from one to an adjacent filament,
it is discontinuous and its slope changes sign [Fig. \ref{figura10} (a)].
The intra-filament linearity of \(\Uptheta ({\rm x})\) differs from 
the sinusoidal expression
%\(\Uptheta ({\rm x}) = \Uptheta_0 \sin (2\uppi \, {\rm x}/{\cal P}) \) 
that was guessed for 
a ``splay-bend'' nematic phase \cite[(c)]{meyerdozov}.
The inter-filament discontinuity reflects
the local orientationally nematic character of the filamentary phase 
in the inter-filament regions,
while in the intra-filament regions 
the filamentary phase has a local orientationally polar character,
with the change of sign that reflects the polarity alternation 
between two consecutive filaments.
On moving from the center of a filament to its border,
the ordinate of the major peak decreases 
while that of the minor peak increases 
until the two ordinates become (essentially) equal at the border; 
this variation in the ordinate of the two peaks reverts 
in the adjacent filament,
consistently with the polarity alternation 
between two consecutive filaments [Fig. \ref{figura10} (b)]. 
Following the statistical-physics custom,
the periodicity of \(f \left( \upvarphi \left| {\rm x} \right. \right)\) along the \textsl{x} axis 
can be summarized by 
the periodicity of the polar order parameter 
\begin{equation}
S_{1,{\hat {\mathbf p}}}({\rm x}) = \int_0^{2\uppi} d\upvarphi \,
{\hat{\mathbf u}}(\upvarphi) \,\cdot\, {\hat{\mathbf p}}({\rm x}) 
f \left( \upvarphi \left| {\rm x} \right. \right)
\label{eqS1p}
\end{equation}  
and the periodicity of the nematic order parameter
\begin{equation}
S_{2,{\hat{\mathbf p}}}({\rm x}) = \int_0^{2\uppi} d\upvarphi \,
\left[ 2 \left( 
{\hat{\mathbf u}}(\upvarphi) \,\cdot\, {\hat{\mathbf p}}({\rm x}) 
\right)^2-1\right]
f \left( \upvarphi \left| {\rm x} \right. \right) \,.
\label{eqS2p}
\end{equation}  
The values of \(S_{1,{\hat{\mathbf p}}}({\rm x})\) are large
in the intra-filament regions and 
abruptly decrease to essentially zero
in the inter-filament regions (Fig. \ref{figura11}).
The values of \(S_{2,{\hat{\mathbf p}}}({\rm x})\) too are large
in the intra-filament regions and
remain as such 
in the inter-filament regions,
although they too experience a little decrease
at the border of two adjacent filaments (Fig. \ref{figura11}).
In addition to Eq. \ref{eqS1p} and Eq. \ref{eqS2p},
where the polar orientational order and the nematic orientational order
are assessed with respect to \(\hat{\mathbf p} ({\rm x}) \), i.e. the local polar director,
one can also calculate the polar order parameter and
the nematic order parameter with respect to 
the \textsl{y} axis, i.e., the axis parallel to the axes of the filaments:
\begin{equation}
S_{1,{\rm {y}}} ({\rm x}) = \int_0^{2\uppi} d\upvarphi \, 
%\cos \upvarphi 
{\hat{\mathbf u}}(\upvarphi) \,\cdot\, {\hat{\mathbf y}} 
f \left( \upvarphi \left| {\rm x} \right. \right) \,; 
\label{eqS1y}
\end{equation}
\begin{equation}
S_{2,{\rm {y}}} ({\rm x}) = \int_0^{2\uppi} d\upvarphi \, 
%[2\cos^2 \upvarphi-1] 
[ 2 ({\hat{\mathbf u}}(\upvarphi) \,\cdot\, {\hat{\mathbf y}})^2 - 1]
f \left( \upvarphi \left| {\rm x} \right. \right) \,. 
\label{eqS2y}
\end{equation}
Glancing at the  graph of \( S_{1,{\rm {y}}} ({\rm x})\) is another way 
to appreciate the polarity alternation of two consecutive filaments.
Recognizing that
the value of \( S_{1,{\rm {y}}}  \) at 
\({\rm x} = (2j+1)R \;, j\in {\mathbb Z}\),  is equal to 0
\begin{figure}
\includegraphics{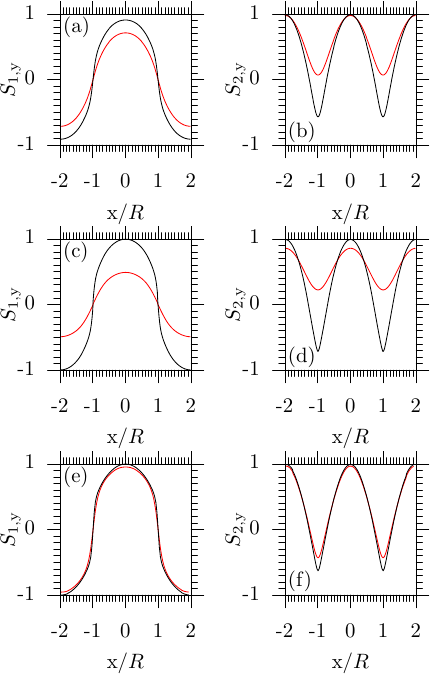}
\caption{
Examples of the graph of
\(S_{1,{\rm y}}({\rm x})\)  and \(S_{2,{\rm y}}({\rm x})\)
in
the filamentary phase of a system of hard minor circular arcs 
from second-virial density-functional theory at values
of \(\uprho^{\star} = \uprho \ell^2\), with \(\ell\) 
the length of a circular arc,
close to, and far from, the transition to the isotropic phase or the nematic phase.
Specifically, a system of hard circular arcs with:
(a,b) \(\uptheta=0.5\) at values
of \(\uprho^{\star} = 19.2\) 
(red or gray) and 
\(\uprho^{\star} =  24.3\) (black);
(c,d) \(\uptheta=1\) at values 
of \(\uprho^{\star} =  10.4\) (red or gray) and
\(\uprho^{\star} =  24.7\) (black); 
(e,f) \(\uptheta=1.8\) at values
of \(\uprho^{\star} =  15\) (red or gray) and
\(\uprho^{\star} =  24.5\) (black).
}
\label{figura12}
\end{figure}
and the 
value of \( S_{2,{\rm {y}}}  \) at these particular values of \( {\rm x}\)
is different from 0 is 
another way to appreciate the local orientationally nematic character
of the filamentary phase at the border of two consecutive filaments (Fig. \ref{figura12}).
While \(\left \langle S_{1,{\rm {y}}}  \right \rangle\) is naturally equal to 0 as long as filaments alternate polarity, 
\(\mathtt{S}_2 = \left \langle S_{2,{\rm {y}}}  \right \rangle\) 
is different from 0 and
its values as a function of \(\uprho\) 
%for a system of hard circular arcs with \(\uptheta = 0.5\)
%compare reasonably well 
can be compared with the
corresponding values that 
were obtained in the Monte Carlo numerical simulations \cite{arcodue} (Fig. \ref{figura13}).

\subsection{mechanism of ``diffusion'': Monte Carlo numerical simulation}
\label{nonema2}
\begin{figure*}
\includegraphics[scale=1.5]{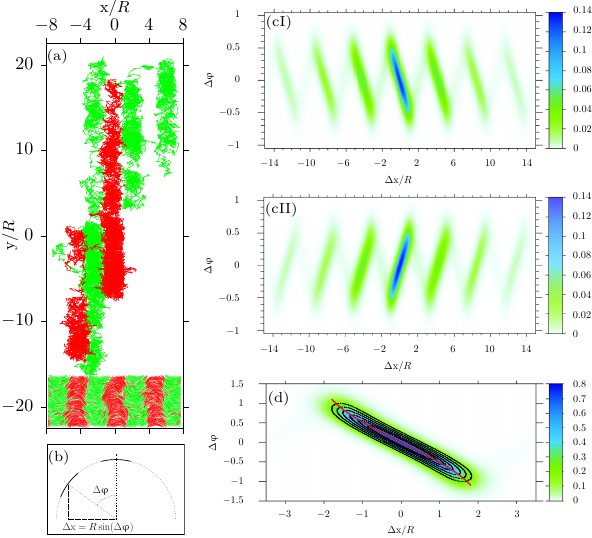}
\caption{
(a)  Example trajectories of hard minor circular arcs 
whose orientation is 
either accordant with the dominant polarity of the filaments 
that has been arbitrarily considered as positive (red or darker gray) 
or accordant with the dominant polarity of the filaments 
that has been arbitrarily considered as negative (green or lighter gray). 
They have been taken from a \(10^7\) MC-cycle--long canonical Monte Carlo numerical simulation 
of the filamentary phase  
of a system of hard circular arcs with \(\uptheta = 0.5\) 
at \(\uprho^{\star} = 19.86\).
In the bottom of this panel,
there is a configuration of this system with 
either the hard minor circular arcs in red (darker gray) or green (lighter gray) 
accordingly to their orientation.
(b) Schematic of the variation of \(\Updelta \upvarphi\) with \(\Updelta {\rm x}\) 
in the ideal motion of a hard minor circular arc within a filament.
(c) Contour plot of the histogram 
\({\rm H}_{\uparrow} (\Updelta {\rm x}, \Updelta \upvarphi)\) (I) and
contour plot of the histogram 
\({\rm H}_{\downarrow} (\Updelta {\rm x}, \Updelta \upvarphi)\) (II)
in the filamentary phase at \(\uprho^{\star} = 19.86\).
(d) Contour plot of the histogram 
\({\rm H}_{\uparrow} (\Updelta {\rm x}, \Updelta \upvarphi)\)
at \(\uprho^{\star} = 30.37\)
and level curves (black) and
graph of the function \(\Updelta \upvarphi = \arcsin \left( -  \Updelta {\rm x} / R\right) \) (red or gray).
}
\label{figura14}
\end{figure*}
%In (soft) condensed-matter physics and statistical physics, 
By the (importance sampling) Monte Carlo method \cite{MCorigin,MCWood,MClibri},
one can calculate thermodynamical and structural
but,
since this method is devoid of any reference to the factual time,
not dynamical properties of a N-particle system.
In any calculation, the Monte Carlo method anyway produces a sequence of configurations.
This sequence of configurations can be ordered, i.e., 
it becomes a succession of configurations.
One can presuppose that two consecutive subsuccessions of configurations,
each one with the same number of configurations,
are also separated by the same fictitious ``time''.
In any calculation, the molecular dynamics method \cite{MClibri} also produces 
a sequence of configurations.
This sequence is naturally ordered, i.e., 
the succession of configurations is the same (within numerical accuracy)
that nature would produce.
Two consecutive subsuccessions of configurations,
each one with the same number of configurations,
are naturally separated by the same physical time.
One can presuppose that the succession of configurations 
that the Monte Carlo method produces 
does not differ too much from
the succession of configurations 
that the molecular dynamics method would produce.
In the Monte Carlo method, a subsuccession is usually taken as a Monte Carlo cycle,
i.e., a number of consecutive configurations that, essentially, coincide with N,
to pretend that in the Monte Carlo method 
the entire N-particle system has moved as it would have moved in the molecular dynamics method, 
where the entire N-particle system moves every time step, and 
the ``time'' that separates two consecutive Monte Carlo cycles 
can be denominated as the ``Monte Carlo time''.
In either method, 
one can imagine to suitably project the succession of configurations 
so as 
to obtain a graph of the trajectory of the N-particle system.
In either method, 
one can also choose a representative particle and 
repeat the suitable projection of the succession of its positions 
so as 
to obtain a graph of the trajectory of this representative particle.
%These operations of projection attenuate the difference between the Monte Carlo method and the molecular dynamics method.
If all the above presuppositions and pretence about the Monte Carlo method are reasonably met
then one can hope that 
the trajectories 
that the Monte Carlo method produces
imitate 
the natural trajectories 
that the molecular dynamics method would produce.
If so, the Monte Carlo method can provide an ``impression'' 
of the mechanism of diffusion that is operative in a N-particle system.

With this, necessary, premise,
the mobility 
in the filamentary phase of a system of \({\rm N} = 5400\) hard circular arcs with \(\uptheta = 0.5\) 
was investigated by 
the Monte Carlo method in the canonical statistical ensemble \cite{MCorigin,MClibri} 
at four values of \(\uprho\):
\(\uprho^{\star} = \uprho \ell^2 =\) 
\(19.86 \); 
\(25.47 \);
\(30.37 \);
\(36.91 \);
with \(\ell\) the length of a circular arc.
The configurations
with which the present four Monte Carlo numerical simulations were initiated
were obtained in 
the previous numerical simulations 
that were conducted by 
the Monte Carlo method in the isobaric(-isothermal) statistical ensemble \cite{MCWood,MClibri} 
at the respective values of pressure:
\(P^{\star} =\upbeta  P \ell^2 =\)
\(45 \); 
\(55 \);
\(60 \);
\(80 \) \cite{arcodue}.
The present four Monte Carlo numerical simulations were conducted
by the same computational protocol of the previous Monte Carlo numerical simulations \cite{arcodue}
except for the absence of any attempt to modify the container.

The motion 
of a hard minor circular arc in the filamentary phase
is connected to the polarity alternation of the filaments.
If the orientation 
of a hard minor circular arc 
is accordant with
the dominant polarity of the filament in which it lives 
then it can move relatively easily and stay for long within this filament. 
If the orientation 
of a hard minor circular arc 
is discordant with
the dominant polarity of the filament in which it lives
then it moves clumsily within this filament and
tends to abandon rapidly this filament
towards an adjacent filament 
whose dominant polarity is accordant with its orientation.
Thus, a hard minor circular arc tends to 
preferably translo-rotationally move within a filament 
that shares the same polarity whereas,
if it succeeds to intrude into an adjacent filament with opposite polarity,
it tends to abruptly 
either advance to the successive filament or retreat to the original filament
[Fig. \ref{figura14} (a)].
 
%While within a filament whose dominant polarity is accordant with its orientation and
%has been considered either positive or negative,
%a hard minor circular arc ideally tends to move 
%as if it is adapting to the parent (semi)circumference from which it was severed.
Within a filament a hard minor circular arc tends to move as if 
it is adapting to the parent (semi)circumference from which it was severed;
especially if its orientation is accordant with
the dominant polarity of the filament. 
This translo-rotational motion is such that 
the translation  transverse to the axis of the filament by a distance \(\Updelta {\rm x}\)
is accompanied by a rotation by an angle
\(\Updelta \upvarphi = \arcsin \left(\mp \, \Updelta {\rm x} / R\right) \) [Fig. \ref{figura14} (b)]. 

For a long succession of configurations,
one can register the linear displacement \(\Updelta {\rm x} (t_{\rm MC}) = 
{\rm x}(t_{\rm MC}) - {\rm x}(t_{\rm MC,0})\) 
and the angular displacement \(\Updelta \upvarphi (t_{\rm MC}) = 
\upvarphi (t_{\rm MC}) - {\rm x}(t_{\rm MC,0})\)  
that any hard minor circular arc makes at an ``instant'' of the ``Monte Carlo time'' \(t_{\rm MC}\) 
with respect to that particular ``instant'' of the ``Monte Carlo time'' \(t_{\rm MC,0}\)
in which it happened 
to be located at the center of a filament and 
%pointed as the axis of this filament.
%One can calculate 
%the histogram, \({\rm H}_{\uparrow} (\Updelta {\rm x}, \Updelta \upvarphi)\),
%for those filaments whose dominant polarity has been arbitrarily considered as positive, and 
%the histogram, \({\rm H}_{\downarrow} (\Updelta {\rm x}, \Updelta \upvarphi)\),
%for those filaments whose dominant polarity has been considered as negative.
pointed according to the dominant polarity of that filament. 
%A further observation is that 
This accordance between the orientation of each hard minor circular arc and 
the dominant polarity of the filament in which it was located at $t_{\rm MC,0}$ 
is conserved throughout all the configurations of the succession: 
e.g., the color of each hard minor circular arc in the configuration at the bottom  
of Fig. \ref{figura14} (a) remains the same in any other configuration of the succession. 
In this way, one can calculate the histogram
$\rm H_{\uparrow}(\Updelta x , \Updelta \upvarphi)$, for those hard minor
circular arcs whose orientation is accordant with the filaments whose dominant
polarity has been arbitrarily considered as positive, and the histogram,
$\rm H_{\downarrow}(\Updelta x , \Updelta \upvarphi)$, for those hard minor
circular arcs whose orientation is accordant with the filaments whose dominant
polarity has been arbitrarily considered as negative.
The two histograms are specular. 
Their contour plots feature a sequence of more intense oblique primary stripes
which are intercalated by significantly less intense specularly oblique secondary stripes
[Fig. \ref{figura14} (c)].
In the case of  \({\rm H}_{\uparrow(\downarrow)} (\Updelta {\rm x}, \Updelta \upvarphi)\)
the primary stripes are inclined left(right)wards and 
their centroids are approximately separated by \(4R\) 
[Fig. \ref{figura14} (c)].
In fact, the slope of the inclination is essentially determined by 
the slope of 
the function \(\Updelta \upvarphi = \arcsin \left(\mp \Updelta {\rm x} / R\right) \)
calculated at \(\Updelta {\rm x} = 0 \)
whose absolute value is equal to \(\ell/R = \uptheta = 0.5\) [Fig. \ref{figura14} (d)].
The primary stripes correspond to the motion of the hard circular arcs 
within filaments whose dominant polarity is accordant with their orientation. 
The central primary stripe is the most intense; 
it corresponds to the motion of the arcs within the filament in which they were located at $t_{\rm MC,0}$. 
The other primary stripes are
progressively less intense as they progressively depart from the central primary stripe;
%The primary stripes correspond to 
%the motion of the hard minor circular arcs 
%whose orientation is accordant with 
%the dominant polarity of the filament at \(t_{\rm MC,0}\). 
%The central primary stripe is the most intense; 
%i.e., the hard minor circular arcs tend to stay for long 
%within a filament whose dominant polarity is of the same sign as their orientation.
%The other primary stripes are progressively less intense 
%as they progressively depart from the central primary stripe;
they correspond to the motion of those hard minor circular arcs 
that had succeeded to abruptly move to successive filaments 
whose dominant polarity is of the same sign as their orientation.
The secondary stripes correspond to 
the motion of those few hard minor circular arcs 
that had been captured 
while in a filament whose dominant polarity
is discordant to their orientation.

The mechanism of mobility 
in the filamentary phase of hard minor circular arcs 
resembles 
the mechanism of diffusion in a smectic phase 
of (hard) elongate particles \cite{diffuSAesp,diffuSAsim}:
the intra-filament regions in the filamentary phase act for 
a hard minor circular arc that is orientated accordingly to 
the filament dominant polarity 
as the intra-layer regions in a smectic  phase act for 
a (hard) elongate particle,
with both types of particle domiciling in the respective preferable regions;
the intra-filament regions in the filamentary phase act for 
a hard minor circular arc that is orientated discordingly to 
the filament dominant polarity 
as the inter-layer regions in a smectic phase act for a (hard) elongate particle,
with both types of particle escaping from the respective unpreferable regions.

This mechanism of mobility is another sign of 
the non-nematicity of the filamentary phase.

\section{conclusion}
\label{conclusione}
There are two conclusions that can be drawn from the results of Section \ref{nonema}. 
The primary conclusion concerns the objective of the present article,
namely, the provision of further results 
in support of 
the non-nematicity of the filamentary phase. 
The secondary conclusion concerns a supplement of the present article, 
namely, the comparison between 
the second-virial density-functional theory and Monte Carlo numerical simulations.
 
Both the second-virial density-functional theory results and the 
present Monte Carlo numerical simulation results confirm 
the conclusion that was drawn from 
the previous Monte Carlo numerical simulation results \cite{arcodue}:
the filamentary phase is not nematic but smectic-like or, more precisely \cite{notuccia}, columnar-like.
Consistently with the form of the positional two-particle correlation functions 
that were previously calculated \cite{arcodue},
the form of the positional one-particle density functions 
that have been presently calculated 
is undulate.
Consistently with this structure of the filamentary phase, 
%the present Monte Carlo numerical simulations find that 
a hard minor circular arc 
tends to stay for long within a filament whose dominant polarity is of the same sign of its orientation
%suitably translo--rotates within a filament
%if its orientation is accordant with the dominant polarity of that filament
whereas, 
if it succeeds to intrude into an adjacent filament whose dominant polarity is of
the opposite sign of its orientation,
it abruptly moves 
either forward to the successive filament 
or backward to the original filament,
both these filaments having the same dominant polarity:
a mechanism of mobility reminiscent of 
the mechanism of diffusion that is operative in a smectic phase of (hard) elongate particles 
in three dimensions \cite{diffuSAesp,diffuSAsim},
with the filaments of the same (opposite) dominant polarity 
that act for the hard minor circular arc in the filamentary phase 
as  the intra(inter)-layer regions act for a (hard) elongate particle in the smectic phase in three dimensions.
Thus, the filamentary phase should not be confounded with a ``modulated'' [``splay(-bend)''] nematic phase.

This conclusion, 
that derives from the results on systems of hard minor circular arcs,
can presently apply to systems of these hard curved particles.
In systems of other (hard, curved) particles,
the formation of a ``modulated'' [``splay(-bend''], truly nematic, phase
cannot be presently excluded.
%Specific investigations should be conducted in this respect.
Yet, the results on systems of hard minor circular arcs 
already suffice to raise the doubt as to 
whether such a ``modulated'' [``splay(-bend)''], truly nematic, phase could ever exist.
To qualify for being nematic a phase must be positionally uniform.
If a phase is ``modulated'' nematic, 
the particles that form this phase experience 
a suitable translo-rotational motion 
that can be ``smoothly'' accomplished 
so that a (mass, number) density non-uniformity does not develop.
Examples of such translo-rotational motions are 
the translo-rotational motion 
that a chiral particle executes in a cholesteric phase
\cite{colesterici2,colesterici3}
%\cite{colesterici1,colesterici2,colesterici3} 
and 
the translo-rotational motion 
that a (hard) helical particle executes in a screw-like nematic phase 
\cite{nemavite1,nemavite2,nemavite3}.
It is not clear what shape and size (hard) particles should have
so as to create a (polar, nematic) director profile
such as the one that was guessed for a ``splay-bend'' nematic phase \cite[(c)]{meyerdozov}
and ``smoothly'' move along it 
without the development of a ``(mass, number) density wave''.
Or, instead, the creation of such a (polar, nematic) director profile
has to be necessarily associated to 
the development of  a ``(mass, number) density wave''.
These considerations, 
that derive from  
the results on systems of hard minor circular arcs  in the two--dimensional Euclidean space \({\mathbb R}^2\),
should be applicable to 
not only other systems of (hard) particles in the two--dimensional Euclidean space \({\mathbb R}^2\) 
but also  other systems of (hard) particles in the three--dimensional Euclidean space \({\mathbb R}^3\),
as the results on systems of hard spherical caps in the three--dimensional Euclidean space \({\mathbb R}^3\)
already indicated \cite{notuccia}.

Irrespective of the dimensionality of the (non-)Euclidean space,
any claim of having dealt with or found 
a ``modulated'' nematic phase, in general, 
or a  ``splay(-bend)'' nematic phase, in particular, 
cannot naturally prescind from,
theoretically, allowing a positional non-uniform phase to exist in the calculations and,
experimentally, scrutinizing the positional uniformity of 
the actual phase under investigation.
%it is a matter of basic rigor and common sense.

In confirming the non-nematicity of the filamentary phase,
use has been made of the second-virial density-functional theory.
Even though it was not the objective of the present article,
a comparison of the results of the second-virial density-functional theory
to the results of Monte Carlo numerical simulations is pertinent
as well as pertinent is a reflection on this comparison.

The second-virial density-functional theory has 
the merit of reproducing the filamentary phase 
that the Monte Carlo numerical simulations revealed \cite{arcodue}.
However, the sequence of the phases and the equations of state,
as well as the one-particle density functions and nematic order parameters, 
that it produces 
compare reasonably well 
with the corresponding results 
from Monte Carlo numerical simulations
only for systems of hard circular arcs with \(\uptheta  = \, ({\rm presumably}\, \leq) \, 0.5\). 
%\whatzit{}
In a way paradoxically, the principal problem 
that the second-virial density-functional theory encounters
is not the reproduction of the filamentary phase at high density,
which it actually succeeds to reproduce, but 
the production of an excessively thermodynamically stable nematic phase
and, especially, the incapability to deal with 
the no ordinary clustering in the higher-density isotropic phase
which is revealed by the straightening of the equation of state of the isotropic phase
that was observed in the Monte Carlo numerical simulations \cite{arcodue, notuccia}.
The consequence of such incapability is 
that the thermodynamic stability of the isotropic phase is depressed so much 
that the nematic phase is excessively favored and
the filamentary phase in systems of hard circular arcs with \(\uptheta \geq 1\) 
occurs at too lower density.
To remedy to this deficiency, 
one way could be to add 
virial terms of higher order 
in the expression 
of the free energy of the isotropic phase.
Even leaving aside the computational difficulty that such an effort requires,
one may doubt whether
a too-early truncated virial series expansion, or even the entire virial series expansion,
could succeed in reproducing the equation of state of the isotropic phase at higher density,
with its strengthening that might be conjectured as
the symptom that the entire virial series expansion have exhausted its convergency. 
Only a rigorous and systematic calculation of the virial coefficients, 
such as those calculations that have been recently accomplished 
for hard convex particles in two dimensions and three dimensions \cite{wagner},
can contribute to clarify that doubt.
To remedy to that deficiency, 
another way could be  to bypass the virial series expansion and 
excogitate  an analytic, albeit heuristic, equation of state for the isotropic phase 
that is accurate also at higher density 
so to arrive at a better approximation of the free energy of the isotropic phase,
along the same line that was followed
for a system of hard infinitesimally--thin discs in three dimensions \cite{dischijpcb}.

\section*{acknowledgements}
The authors acknowledge support from 
the Government of Spain 
under Grant No. FIS2017-86007-C3-1-P.


\begin{thebibliography}{99}
\bibitem{splaybend} 
The generic term ``modulated'' is preferred to 
the particular term ``splay-bend'' 
which is at least confusing.
% if not erroneous.
In two dimensions, 
there cannot be a difference between  
a ``splay'' distorsion and  a ``bend'' distorsion of \(\hat n\):
be \(\hat{\mathbf u}_{par} \) a unit vector 
which has been subjectively chosen to describe the orientation of a particle and 
be \(\hat{\mathbf u}_{per}\) a perpendicular unit vector; 
any ``splay'' variation of \(\hat{\mathbf u}_{par}\) corresponds to 
a ``bend'' variation of \(\hat{\mathbf u}_{per}\) and vice versa:
all depends on the subjective choice of \(\hat{\mathbf u}_{par}\) and 
hence of \(\hat{\mathbf u}_{per}\);
while this holds true for any particle, 
it holds even truer for a curved particle, 
for which, differently from an elongate or flat particle,  
it is admittedly ``difficult'' to 
objectively choose its ``principal'',  long or short, axis.
\bibitem{colesterici3}
The authors are not informed of molecular dynamics numerical simulation investigations 
that focus on the mechanism of diffusion of chiral particles in a cholesteric phase,
although such a mechanism of diffusion is envisageable,
with a chiral particle that, 
while it translates along, it also suitably rotates round the cholesteric axis.
\end{thebibliography}
\end{document}